\documentstyle{epsf}
\documentstyle[iopfts]{ioplppt}

\begin{document}
\title{The geometry of the Barbour-Bertotti theories
II. The three body problem}
\author{L\'{a}szl\'{o} \'{A} Gergely \dag \ddag\ and Mitchell McKain \S}
\address{\dag Laboratoire de Physique Th\'{e}orique,
Universit\'{e} Louis Pasteur
3-5 rue de l'Universit\'{e} 67084 Strasbourg, France}
\address{\ddag KFKI Research Institute for Particle and Nuclear Physics,
Budapest 114, P.O.Box 49, H-1525 Hungary}
\address{\S Department of Physics, University of Utah,
Salt Lake City, Utah 84112}
 
\begin{abstract}
We present a geometric approach to the three-body problem 
in the non-relativistic context of the Barbour-Bertotti theories. 
The Riemannian metric characterizing the dynamics is analyzed in 
detail in terms of the relative separations. Consequences of a 
conformal symmetry are exploited and the sectional curvatures of 
geometrically preferred surfaces are computed. The geodesic motions 
are integrated. Line configurations, which lead to curvature 
singularities for $N\neq 3$, are investigated. 
None of the independent scalars formed from the metric and curvature 
tensor diverges there. 

\end{abstract}
 
\section{Introduction}
  
The nonrelativistic dynamical models of Barbour and Bertotti 
\cite{BB1,BB2} arose from the criticism of the concepts 
of absolute space and time. They describe a classical interacting 
$N-$particle system subjected to the Hamiltonian, momenta and 
angular momenta constraints. The invariance group of the theory 
is the Leibniz group \cite{BB1}, which includes time-dependent 
translations and rotations together with the monotonous but otherwise 
arbitrary redefinition of time.

In a previous paper \cite{Gergely} being referred hereafter as 
paper {\bf I}, one of the present authors has analyzed the 
underlying geometry of the Barbour-Bertotti theories. The reduction 
process on the Lagrangian was carried out by solving the constraints, 
arriving to a Riemannian line element. This reduction was possible 
for all configurations but the line ones. The emerging Riemannian metric 
$G$ was shown to represent the first fundamental form of the orbit space 
of the Leibniz group. The geodesics in this metric characterize the free
motions, pertinent to constant potential $V$. For a generic potential
$V\neq const$ the motions are geodesics of the conformally scaled 
Jacobi metric $-2V(x)G$. 

In {\bf I} the Riemann tensor and curvature scalar were computed 
in terms of the vorticity tensors of the generators of rotations. Then 
the curvature scalar was expressed in terms of the principal moments of 
inertia and the number of particles. An analysis based on this expression 
allowed us to conclude that the line configurations represent curvature 
singularities for $N\neq 3$. One would like to say more about these 
configurations for the exceptional case $N=3$. This is one of the 
motivations of the present work. 

The second motivation for specializing to three particles is the 
remarkable coincidence between the number of relative separations 
$N(N-1)/2$ among the particles and the dimension $3N-6$ of the reduced 
space
\footnote{$3N-6=N(N-1)/2$ holds also for $N=4$. A discussion of the 
adequate coordinates for this case can be found in \cite{LR2}.} 
. This feature enables one to employ the distances as a symmetric 
set of variables of the space of orbits and to analyze in detail the 
underlying Riemannian geometry. 

We discuss and picture the space of orbits in Sec. 2.  
Similar discussions can be found in \cite{Barbour2} and \cite{LR}. 
The space of orbits being a manifold with boundary, we announce 
and prove the conditions a geodesic reaching this boundary has to obey.

The rest of the paper is organized as follows. First we particularize 
the generic expression of the curvature scalar obtained in {\bf I} to 
the case of three particles. For this purpose we compute in Sec. 3 the 
explicit expressions of the principal moments of inertia in terms of 
the relative separations. We find that the curvature scalar reduces to 
a particularly simple form in the case of three particles. 
 
We reveal more details of the geometry in Sec. 4. There we give the 
reduced metric on the space of relative separations first in terms of 
the distances and second in terms of a radial coordinate and suitably 
chosen angular variables. Using the second set of variables the Riemann 
and Ricci tensors are computed. None of the scalars formed from the 
metric and the Riemann tensor diverges in the line configurations. 

We show that the metric is conformally flat. Furthermore, the metric 
has a conformal symmetry. This enables us to compute the extrinsic 
curvature of the ellipsoid surfaces orthogonal to the conformal Killing 
vector in Sec. 5 and to demonstrate in Sec. 6 that the radial lines are 
geodesics. 

The purpose of Sec. 5 is to analyze the eigenvalue problem for the 
Ricci tensor. Remarkably, one of the Ricci principal directions is the 
conformal Killing vector. The sectional curvature of the ellipsoid 
surfaces orthogonal to the conformal Killing vector is found by the 
Gauss relation. The sectional curvature of the conical surfaces 
orthogonal to the other eigenvectors are also computed. All eigenvalues 
of the Ricci tensor are finite for the line configurations.

Free motions are shown in Sec. 6. to be geodesics of the reduced space. 
The geodesic equation is integrated in the space of distances.

Throughout the paper we use the following notations. Latin indices denote 
components in the space of distances. They run from $1$ to $3$ and are 
raised or lowered with a non-flat metric. Greek indices label quantities 
pertinent to different eigenvectors of the Ricci tensor. There are few 
exceptions under these rules, indicated where necessary in the text. 
Summations are explicitly written whenever the summation convention can not 
be applied. Partial derivatives are denoted by comma.
   
\section{The space of orbits}

The nine coordinates characterizing the positions of the particles can
be chosen as the relative separations $a_{i}=(a,b,c)$, the Euler angles 
$\alpha _{i}=(\alpha ,\beta ,\gamma )$ of the normal to the plane of the
three particles and the coordinates of the center of mass $x^{i}$.  
By freezing the translational and the rotational degrees of freedom,
we fix the coordinates of the center of mass and the Euler angles 
respectively, such that the distances $(a,b,c)$ will coordinatize the 
reduced space. 
Before proceeding with the analysis of the metric and curvature 
properties of the reduced configuration space, we briefly describe and 
picture this space in the present section. 
 
\begin{figure}[tbh]
\epsfysize=7cm
\centerline{\hfill
\epsfbox{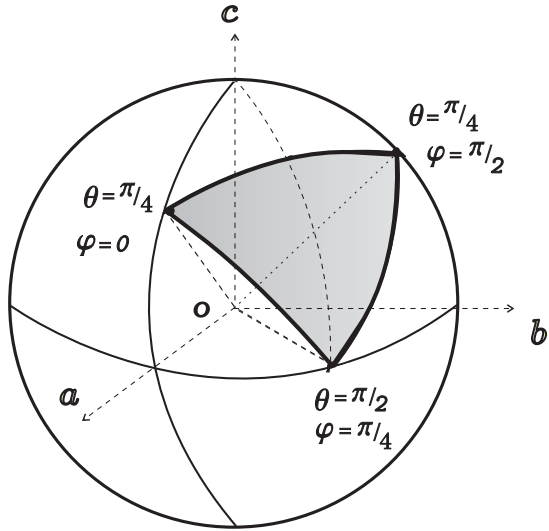}\hfill}
\vskip1cm
\caption{The intersection of an $r=constant$ surface with the space of
orbits for three particles in the Cartesian frame of the distances is a
spherical triangle, the boundaries of which are arcs of great circles. }
\end{figure}

The domain of the polar and azimuthal angles $\theta $ and $\varphi $
defined in the Cartesian system spanned by the coordinate lines ${a_{i}}$ is
restricted first by the positivity of $a_{i}$:
\begin{equation}
0\leq \theta \leq {\frac{\pi }{2}}\ ,\qquad 0\leq \varphi \leq {\frac{\pi }{2
}}\ ,  \label{domain1}
\end{equation}
second by the triangle inequality:
\begin{equation}
\mid \sin \varphi -\cos \varphi \mid \leq \cot \theta \leq \sin \varphi
+\cos \varphi \ .  \label{domain}
\end{equation}
Therefore the reduced space can be imagined like a diverging beam of rays
starting from the origin, with a triangular section, as can be seen on Fig 1.
The boundaries of the admissible domain of the reduced space are defined by
the equalities in Eq. (\ref{domain}) as planes passing through the origin.
In consequence each constant radius section of the reduced space is bounded 
by arcs of great circles. The origin $O$ corresponds to the unphysical
situation of all three particles in the same location while the planar
boundaries to the situation of the three particles on a line. The edges of
the beam (the intersection of two plane boundaries) are again unphysical:
there the positions of two of the three particles coincide. On all
boundaries of the space of orbits the inertia matrix has vanishing 
determinant, consequently the metric (32) of {\bf I} is ill-defined there. 
On the other hand the scalar curvature is well behaved in the line 
configurations, suggesting that the boundary planes (including the edges 
but not the origin $O$) are merely coordinate singularities. 
Topologically the region inside the boundaries is the quotient space 
$R^9/E(3)$, where $E(3)=R^3\rtimes SO(3)$ 
and is homeomorphic to half of $R^3$.
The boundary is homeomorphic to $R^2$ \cite{LR2}.

The special situation of the restricted three-body problem, with a lighter
third body orbiting about the other two heavier, whose separation is not
changing is confined within intersections of the space of orbits (the
''beam'') with one of the planes $a_{i}=const$.

As physically the line configurations are no more special then others, 
our description has to allow the geodesics to reach and to leave the 
boundaries. We then require that a geodesic reaching the boundary at 
some instant $t_0$ should not leave the admissible domain of the 
$\{a_i\}$ space, by imposing the following boundary conditions:
\begin{eqnarray}
{\bf \dot a}_{\perp}(t_0)=0 \ ,
\label{refl1}
\\
{\bf \ddot a}_{\perp}(t_0)<0 \ .
\label{refl2}
\end{eqnarray}
Here ${\bf \dot a}_{\perp}$ and ${\bf \ddot a}_{\perp}$ represent the
projections perpendicular to the boundary 
of the first and second derivatives of the position vector 
${\bf a}=(a,b,c)$ with respect to the geodesic parameter. 
The perpendicularity is meant in an Euclidean sense, by projecting 
${\bf \dot a}$ and ${\bf \ddot a}$ to the {\it outer} normal to the 
boundary surface.

{\it Proof of the boundary conditions:}

It is well known that the motion of a three body system with vanishing 
angular momentum ${\bf L}=\sum_A {\bf L}_A=0$ 
\footnote{Here and later in Eqs. (\ref{distCM}) and (\ref{sum2}) we use the 
conventions of {\bf I}: capital Latin indices count the particles and lower 
case Latin indices their coordinates. Summation is explicitely indicated 
only over particles.}
is confined to a plane \cite{Whittaker}. 
By using Cartesian coordinates $(x,\ y,\ z)$ we choose 
$z_A=0=\dot z_A$ and then the constraints 
$\sum_A {\bf P}_{Az}=\sum_A {\bf L}_{Ax}=\sum_A {\bf L}_{Ay}=0$ 
are trivially satisfied.
The distances ${a_i}$ are given as:
\begin{eqnarray}
a^2=(x_2-x_3)^2+(y_2-y_3)^2\ ,\nonumber \\
b^2=(x_3-x_1)^2+(y_3-y_1)^2\ ,\nonumber \\
c^2=(x_1-x_2)^2+(y_1-y_2)^2\ .\label{dist}
\end{eqnarray}
As parameter of the geodesic the time can be chosen. The time derivative 
of (\ref{dist}) gives $\{\dot a_i\}$ 
\begin{eqnarray}
a\dot a=(x_2-x_3)(\dot x_2-\dot x_3)+(y_2-y_3)(\dot y_2-\dot y_3)
\ ,\nonumber \\
b\dot b=(x_3-x_1)(\dot x_3-\dot x_1)+(y_3-y_1)(\dot y_3-\dot y_1)
\ ,\nonumber \\
c\dot c=(x_1-x_2)(\dot x_1-\dot x_2)+(y_1-y_2)(\dot y_1-\dot y_2)
\ .\label{ddist}
\end{eqnarray}
From among the "velocities" 
$\dot x_1,\ \dot x_2,\ \dot x_3,\ \dot y_1,\ \dot y_2,\ \dot y_3$
only three are independent due to the remaining constraints 
$\sum_A {\bf P}_{Ax}=\sum_A {\bf P}_{Ay}=\sum_A {\bf L}_{Az}=0$. 
However for the purpose of the proof it is not necessary to make 
this manifest. We just comment that in general $\{\dot a_i\}$ depend 
on three arbitrary velocities and in the majority of cases they are 
independent. A notable exception is given by the collinear 
configurations, as will be seen later. 
In the computation of 
the second derivatives we employ that the motions are free, 
$\ddot x_A=\ddot y_A=0$: 
\begin{eqnarray}  
a\ddot a+{\dot a}^2=(\dot x_2-\dot x_3)^2+(\dot y_2-\dot y_3)^2
\ ,\nonumber \\
b\ddot b+{\dot b}^2=(\dot x_3-\dot x_1)^2+(\dot y_3-\dot y_1)^2
\ ,\nonumber \\
c\ddot c+{\dot c}^2=(\dot x_1-\dot x_2)^2+(\dot y_1-\dot y_2)^2
\ .\label{dddist}
\end{eqnarray}
Then we study the collinear configurations. By a rotation in the 
$xy$ plane, it can be achieved that the collinearity arises 
on the $x$-axis. We label the particles as $(3,\ 1,\ 2)$ 
in order of increasing $x$-coordinate. In this way 
$x_3-x_1=-b,\ x_1-x_2=-c,\ x_2-x_3=a$ and the boundary is 
expressed by $a=b+c$, with the outer normal 
${\bf n}=(1,\ -1,\ -1)$. 
The system (\ref{ddist}) becomes
\begin{equation}
\dot a= (\dot x_2-\dot x_3)\ ,\quad
\dot b=-(\dot x_3-\dot x_1)\ ,\quad
\dot c=-(\dot x_1-\dot x_2)\ .\label{ddistcol}
\end{equation}
It is immediate to show 
${\bf \dot a}_{\perp}={\bf \dot a}\cdot {\bf n}=0$.
Employing Eqs. (\ref{ddistcol}), the system (\ref{dddist}) can be written 
in the form 
\begin{equation}  
(b+c)\ddot a=(\dot y_2-\dot y_3)^2
\ ,\quad
    b\ddot b=(\dot y_3-\dot y_1)^2
\ ,\quad
    c\ddot c=(\dot y_1-\dot y_2)^2
\ .\label{dddistcol}
\end{equation}
By simple algebra we find then 
\begin{equation}
{\bf \ddot a}_{\perp}={\bf \ddot a}\cdot {\bf n}
=-\frac{[b(\dot y_1-\dot y_2)+c(\dot y_3-\dot y_1)]^2}{bc(b+c)}<0
\end{equation} 
which completes the proof of the boundary conditions, 
Eqs. (\ref{refl1})-(\ref{refl2}).

Q.E.D.

\section{The curvature scalar for three particles}
 
In this section we work out the expression for the curvature scalar
(given in {\bf I} in terms of the principal moments of inertia) for 
the particular case of three particles. For this purpose first we write 
the principal moments of inertia in terms of the relative separations.
 
By choosing the positions of the three particles in some initial 
coordinate system (Fig. 2) originating in the center of mass as 
$\mathop{x}\limits_{\circ }{}^{1a}=(x,0,0),\ 
 \mathop{x}\limits_{\circ }{}^{2a}=(\xi ,y,0)$ and 
$\mathop{x}\limits_{\circ }{}^{3a}=(-x_{3},-y_{3},0)$, 
the $x_{3}$ and $y_{3}$ coordinates are determined by the condition 
$\sum_A m_A \mathop{x}\limits_{\circ }{}^{Aa}=0$: 
\begin{equation}
x_{3}={\frac{m_{1}x+m_{2}\xi }{m_{3}}}\qquad y_{3}={\frac{m_{2}}{m_{3}}}y
\ .\label{distCM}
\end{equation}
The remaining three coordinates $(x,\xi ,y)$ are related to the relative
distances as:
\begin{eqnarray}
m_{3}^{2}a^{2}  = \left[ m_{1}x+(m_{2}+m_{3})\xi \right]
^{2}+(m_{2}+m_{3})^{2}y^{2}  \nonumber \\
m_{3}^{2}b^{2}  = \left[ (m_{1}+m_{3})x+m_{2}\xi \right] ^{2}+m_{2}^{2}y^{2}
\nonumber \\
c^{2}  = x^{2}+y^{2}+\xi ^{2}-2x\xi \ .  \label{coordist}
\end{eqnarray}

\begin{figure}[ht]
\epsfysize=7cm
\centerline{\hfill
\epsfbox{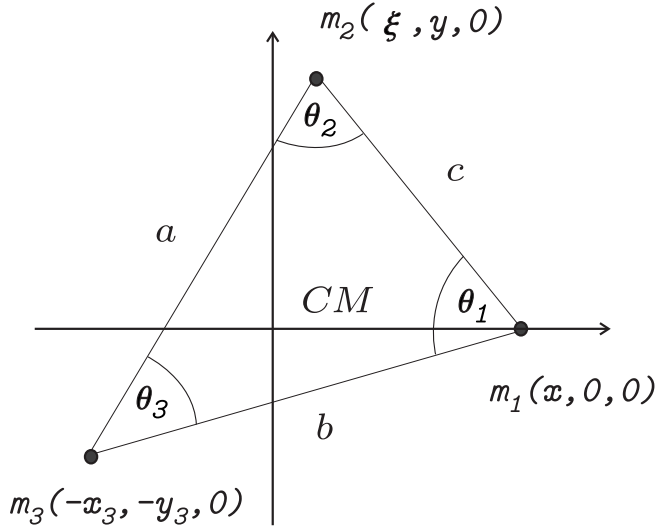}\hfill}
\vskip1cm
\caption{The three particles in the center of mass frame.}
\end{figure}
 
We introduce the positive quantities $A,W$ and $Z$, depending only on
relative distances and with the dimensions of moments of inertia, as
follows:
\begin{eqnarray}
W  = {\frac{m_{1}m_{2}c^{2}+m_{2}m_{3}a^{2}+m_{3}m_{1}b^{2}}{2M}}\ ,\qquad A=
{\frac{m_{2}m_{3}}{m_{2}+m_{3}}}a^{2}\ ,  \nonumber \\
Z^{2}  = {\frac{m_{1}m_{2}m_{3}}{4M}}(a+b+c)(a+b-c)(a-b+c)(-a+b+c)\ .
\label{WZA}
\end{eqnarray}
Heron's formula relates the quantity $Z$ to the area ${\cal A}$ of the
triangle: $Z=2{\cal A}\sqrt{m_{1}m_{2}m_{3}/M}$. From Eqs. (\ref{coordist}) and 
Eqs. (\ref{WZA}) a useful relation is found:
\begin{equation}
\sum_{A}m_{A}\mathop{x}\limits_{\circ }{}^{Ac}\mathop{x}\limits_{\circ
}{}^{Ad}\delta _{cd}=2W  \label{sum2}
\end{equation}

A computation employing the definition of the inertia tensor in the 
center of mass frame (Eq. (21) of {\bf I}) and Eqs. (\ref{coordist}), 
(\ref{WZA}) and (\ref{sum2}) yields the expression of the tensor of 
inertia in terms of distances:
\begin{equation}
\mathop{I}\limits_{\circ }{}_{gh}=
\!\!\left(\!\!
\begin{array}{lll}
I_{11} &  -\sqrt{I_{11}(2W-I_{11})-Z^{2}} &  0 \\
-\sqrt{I_{11}(2W-I_{11})-Z^{2}} &  2W-I_{11} &  0 \\
0 &  0 &  2W
\end{array}
\!\!\right)\!\! 
,\label{intensxxiy}
\end{equation}
where
\begin{equation}
I_{11}={\frac{Z^{2}}{(2W-A)}}\ .  
\end{equation}
From here we find the principal moments of inertia: 
\begin{equation}
I_{1,2}=W\pm \sqrt{W^{2}-Z^{2}}\ ,\qquad I_{3}=2W\ .\label{moments}
\end{equation}
They contain the relative distances and the masses in a symmetric fashion 
($I_{11}$, which is the only term containing the non-symmetric expression $A$, 
dropped out from their expression). By insertion of the principal moments 
of inertia in Eq. (43) of {\bf I} we have an independent check of 
Eq. (\ref{sum2}).

The curvature scalar of the reduced space given by Eq. (51) of {\bf I}, 
written in terms of the distances takes the remarkably simple expression:
\begin{equation}
R={\frac{3}{W}}\ .  \label{R3P}
\end{equation}
As expected from the argumentation in {\bf I}, the curvature scalar is 
well behaved even in the case of all three particles in a line. This 
suggests that in the line configurations a coordinate singularity occurs. 
By contrast, the unphysical situation of all particles in a point 
represents a true curvature singularity, as $W$ vanishes there.
  
\section{Metric properties}
 
In order to find the metric in the space of orbits in terms of the distances
$a_{i}$ we start from the degenerated metric (32) of {\bf I}. Then we pass to the
basis associated with the coordinates $(y^{i},\ \alpha _{j},\ x^{k}),$ where
$y^{i}=(x,\ \xi ,\ y)$. The relevant part of the transformation matrix is
given by
\begin{equation}
\frac{\partial \mathop{x}\limits_{\circ }{}^{Ai}}{\partial y^{j}}=\left(
\delta _{1}^{A}-\frac{m_{1}}{m_{3}}\delta _{3}^{A}\right) \delta
_{1}^{i}\delta _{j}^{1}+\left( \delta _{2}^{A}-\frac{m_{2}}{m_{3}}\delta
_{3}^{A}\right) \left( \delta _{1}^{i}\delta _{j}^{2}+\delta _{2}^{i}\delta
_{j}^{3}\right) \ .
\end{equation}
The $3\times 3$ block with the basis in the 1-forms $dy^{i}$ then is
transformed to the basis $da_{i}$ through the relations (\ref{coordist}).
Finally we get:
\begin{equation}
G^{ij}=
\left(
\matrix{
\displaystyle{\frac{m_{2}+m_{3}}{m_{2}m_{3}}}      &   
\displaystyle{\frac{a^{2}+b^{2}-c^{2}}{2m_{3}ab}}  &   
\displaystyle{\frac{a^{2}-b^{2}+c^{2}}{2m_{2}ac}}  \cr
&&\cr
\displaystyle{\frac{a^{2}+b^{2}-c^{2}}{2m_{3}ab}}  &   
\displaystyle{\frac{m_{1}+m_{3}}{m_{1}m_{3}}}      &
\displaystyle{\frac{-a^{2}+b^{2}+c^{2}}{2m_{1}bc}} \cr
&&\cr
\displaystyle{\frac{a^{2}-b^{2}+c^{2}}{2m_{2}ac}}  &   
\displaystyle{\frac{-a^{2}+b^{2}+c^{2}}{2m_{1}bc}} &   
\displaystyle{\frac{m_{1}+m_{2}}{m_{1}m_{2}}}
}
\right) \ .  \label{3metric}
\end{equation}
The covariant form of the metric has a more complicated expression. 
Eq. (\ref{3metric}) is the induced metric in the space of orbits. It is 
not surprising, that $G^{ij}$ is manifestly symmetric in both masses 
and distances, because the overall translational and rotational degrees 
of freedom have been suppressed.
 
Remarkably, the metric $G^{ij}$ has a conformal symmetry. Indeed, (\ref
{3metric}) is unchanged under multiplying all distances with the same
factor, thus $G^{ij}$ is homogeneous of degree zero in $a_i$. The
Euler theorem for homogeneous functions gives then 
$\sum_{k}G_{\ \,k}^{ij}a_{k}=0$ for any point of the reduced space 
with coordinates $a_{i}$. Then the vector
\begin{equation}
K^{i}=a_{i}=(a,b,c)  \label{Ku}
\end{equation}
is a conformal Killing vector, generating a homothetic motion:
\begin{equation}
{\ell }_{K}\ G^{ij}=-2G^{ij}  \label{Killing}
\end{equation}
The norm and covariant components of this conformal Killing vector 
can be expressed in terms of the previously introduced quantity $W$:
\begin{equation}
K=\sqrt{2W}\ ,\qquad K_{i}={\frac{1}{M}}(m_{2}m_{3}a,\ m_{1}m_{3}b,\
m_{1}m_{2}c)=\nabla _{i}W\ .  \label{Kd}
\end{equation}
Here $\nabla$ denotes the connection compatible with the metric $G$.
 
The distance in the metric $G_{ij}$ of any point from the origin is the
length of the corresponding conformal Killing vector (\ref{Kd}). Thus the
curvature scalar (\ref{R3P}) is just 6 divided by the square of this
distance.
 
Whenever one has a metric $G^{ij}$ with a conformal symmetry, by an
appropriate rescaling a new metric $\bar{G}^{ij}=\Omega ^{2}G^{ij}$ can be
found which has the property, that the conformal Killing vector of $G^{ij}$
is a true Killing vector of the metric $\bar{G}^{ij}$. In our case the
scaling function was found to be $\Omega ^{2}=a^{2}+b^{2}+c^{2}=r^{2}$.
Choosing an adapted coordinate system the metric $\bar{G}^{ij}$ does not
depend on $r$. Such a coordinate system is given by $(ln\ r,\ \theta ,\
\varphi )$ in which the conformal Killing vector of $G^{ij}$ takes the
simple form $K^{i}=(1,0,0)$. Thus the metric $G^{ij}$ is expected to have
the form:
\begin{equation}
G^{ij}=\frac{1}{r^{2}}\cdot \bar{G}^{ij}(angles)\ .
\end{equation}
Unfortunately the attempts to write the metric in terms of the radial
coordinate $r$, supplemented by any two angular coordinates, have resulted
in less symmetric forms.
 
In the search for angular variables which give a reasonably simple form of
the metric, we have found a convenient (but redundant) set given by the
sines
\begin{equation}
s_{i}=\sin \theta _{i}
\end{equation}
of the three angles $\theta _{i}$ (Fig.2.). The condition $\theta
_{1}+\theta _{2}+\theta _{3}=\pi $ constrains the variables $s_{i}$ as
follows:
\begin{equation}
s_{1}^{2}+s_{2}^{2}-s_{3}^{2}=2s_{1}s_{2}\sqrt{1-s_{3}^{2}}
\label{constr1form}
\end{equation}
and all of its cyclic permutations. The square of Eq. (\ref{constr1form})
yields a symmetric form of this constraint:
\begin{equation}
(s_{1}+s_{2}+s_{3})(s_{1}+s_{2}-s_{3})(s_{1}-s_{2}+s_{3})(s_{2}+s_{3}-s_{1})
=4s_{1}^{2}s_{2}^{2}s_{3}^{2}\ .
\label{constr2form}
\end{equation}
From the sine theorem we find that the relative separations are related to
the redundant variables $s_{i}$ as:
\begin{equation}
a_{i}=\frac{r}{S}s_{i}\ ,
\end{equation}
where $S^{2}=s_{1}^{2}+s_{2}^{2}+s_{3}^{2}$. By multiplying Eq. 
(\ref{constr2form}) with $(r/S)^6$ we find $S^2$ in terms of the distances:
\begin{equation}
S^2=\frac{Mr^2Z^2}{m_1m_2m_3a^2b^2c^2}\ .
\label{S2}
\end{equation}
We also introduce the quantity 
\begin{equation}
\sigma ^{2}=\frac{2MWS^{2}}{r^{2}}
           =m_1m_2s_3^2+m_2m_3s_1^2+m_3m_1s_2^2
           =\frac{2M^2WZ^2}{m_1m_2m_3a^2b^2c^2} 
\ , 
\label{sigma2}
\end{equation}
which depends only on the angular variables $s_i$ as can be seen from the 
second form given in Eq. (\ref{sigma2}). The third form arises by inserting
Eq. (\ref{S2}) in the first form and similarly to Eq. (\ref{S2}) is useful 
in rewriting the angular expressions in terms of the distances.
 
By multiplying Eq. (\ref{constr1form}) with $(r/S)^2$ the following relation 
emerges:
\begin{equation}
\sqrt{1-s_3^2}=\frac{a^2+b^2-c^2}{2ab}\ .
\end{equation}
Similar equations follow for the cyclic permutations. Therefore the 
contravariant metric expressed in these new angle variables, while
keeping the basis of the metric in the 1-forms ${da_{i}}$, takes the concise
form:
\begin{equation}
G^{ij}=\frac{\left( M-m_{j}\right) m_{j}}{m_{1}m_{2}m_{3}}\delta
^{ij}+\sum\limits_{k}\frac{\sqrt{1-s_{k}^{2}}}{m_{k}}\mu ^{ijk}\ .
\label{3metricang}
\end{equation}
Here $\mu ^{ijk}$ is the ''square'' of the symbol $\epsilon ^{ijk}$, zero
for any pair of coinciding indices and taking the value $1$ otherwise
and no summation over $j$ applies.
 
The concise form (\ref{3metricang}) of the metric allows us to study its
signature. First we remark that all three diagonal elements of $G^{ij}$ (the
first term in (\ref{3metricang})) are positive. Also all three
sub-determinants of the diagonal elements have similar forms, for example
the sub-determinant of $G^{11}$ is
\begin{equation}
\frac{s_{3}^{2}}{m_{3}^{2}}+\frac{M}{m_{1}m_{2}m_{3}}\ ,
\end{equation}
and are also positive. Finally the determinant $1/G$ of $G^{ij}$,
\begin{equation}
G^{-1}=\frac{M\sigma ^{2}}{m_{1}^{2}m_{2}^{2}m_{3}^{2}}
\end{equation}
is positive except in the singular configuration of particles in a line,
where $\sigma ^{2}$ is zero. Therefore we are in the position to conclude
that the metric is positive definite, modulo the boundaries where it diverges.
 
Direct computation of the curvature scalar from the metric yields
(\ref{R3P}) once again, which in terms of the angular variables
and $r$ takes the form:
\begin{equation}
R=\frac{6MS^2}{\sigma^2r^2}\ .
\label{R3Pang}
\end{equation}
The Ricci tensor can be expressed in terms of the curvature scalar:
\begin{equation}
R_{ij}=R\frac{m_{1}m_{2}m_{3}}{2\sigma ^{2}}\left[ \delta
_{ij}-\sum\limits_{k}\sqrt{1-s_{k}^{2}}\ \mu _{ijk}\right] \ .  \label{Ricci}
\end{equation}
As in three dimensions the Weyl tensor vanishes, the Riemann
tensor is determined completely by the Ricci tensor,
the curvature scalar and the metric:
\begin{equation}
R_{ijkl}=2G_{i[k}R_{l]j}-2G_{j[k}R_{l]i}-RG_{i[k}G_{l]j}\ .
\label{noWeyl}
\end{equation}
Again we write the result of the computation in terms of the
curvature scalar:
\begin{equation}
R_{ijkl}=R\frac{m_{1}^{2}m_{2}^{2}m_{3}^{2}}{2\sigma ^{4}}
\sum\limits_{a}\epsilon_{ija}s_a
\sum\limits_{b}\epsilon_{klb}s_b\ .
\label{Riemann3P}
\end{equation}
All multiplying factors of $R$ in $R_{ijkl}$ and $R_{ij}$ depend only on the
angular variables, thus these tensors depend on $r$ only through
$R$.

Next we study the behavior of the curvature in the line configurations.
Though the curvature scalar is undetermined in the form (\ref{R3Pang}),
it is clearly nonsingular when written in terms of the distances 
(\ref{R3P}), as already remarked. From Eq. (\ref{Riemann3P}) we compute 
the Kretschmann scalar and find that it is nonsingular either:
\begin{equation}
R_{ijkl}R^{ijkl}=R^2\ .
\end{equation}
From Eq. (\ref{noWeyl}) an algebraic relation can be deduced between these
scalars
\begin{equation}
R_{ijkl}R^{ijkl}-4R_{ij}R^{ij}+R^2=0\ ,
\end{equation}
which gives 
\begin{equation}
R_{ij}R^{ij}=\frac{R^2}{2}\ .
\end{equation}
In fact in three dimensions there are only three independent scalars which
can be formed from the Riemann tensor and the metric. These can be given 
either as  
\begin{equation}
R\ ,\qquad R_{ij}R^{ij}\ ,\qquad \frac{Det(R)}{Det(G)}
\label{curvscalars}
\end{equation}
or as the roots of the secular equation $Det(R_{ij}-\lambda G_{ij})=0$
\cite{Weinberg}. To complete the first set (\ref{curvscalars}) we compute
\begin{equation}
\frac{Det(R)}{Det(G)}=0\ ,
\label{scalars}
\end{equation}
while the secular equation will be dealt with in the next section.

Finally we remark that in three dimensions the vanishing of the 
tensor
\begin{equation}
R_{abc}=\nabla_c R_{ab}-\nabla_b R_{ac}
+\frac{1}{4}(G_{ac}\nabla_b R-G_{ab}\nabla_c R)
\end{equation}
is equivalent with the conformal flatness of the metric \cite{Eisenhart}.
The computation gives $R_{abc}=0$, therefore the metric is
conformally flat:
\begin{equation}
G^{ab}=\Phi^2 \eta^{ab}\ .
\end{equation}
Here $\eta_{ab}$ denotes a flat metric. The conformal factor can be
determined by solving the differential equation which arises from
the comparison of the expression (\ref{R3P}) with the formula
\cite{Wald} relating the curvature scalars of the metrics $G^{ab}$
and $\eta^{ab}$ (the curvature of the latter being zero):
\begin{equation}
R=\frac{2G^{ab}\nabla_a\nabla_b\Phi^2}{\Phi^2}
-\frac{3G^{ab}\nabla_a\Phi^2\nabla_b\Phi^2}{2\Phi^4}\ .
\end{equation}
It is easy to prove that
\begin{equation}
\Phi^2=\alpha W\ ,
\label{confac}
\end{equation}
$\alpha$ arbitrary constant, solves the differential equation, by
employing Eqs. (\ref{Kd})
and the relation $\nabla_a K^a=3$, which stems out from the
conformal Killing equation (\ref{Killing}).
 
A coordinate system $\{w_i\}$, related to the Jacobi
coordinates, frequently used \cite{LR} in molecular dynamics, has the
advantageous property that $\eta^{ab}=diag(1,1,1)$. The price one
has to pay in using such coordinates is that they depend on the
distances in a nonsymmetric fashion:
\begin{equation}
w_1=2(B-W)\ ,\quad w_2=2\sqrt{2BW-B^2-Z^2}\ ,\quad w_3=2Z\ ,
\end{equation}
where the notation $B=m_1m_3b^2/(m_1+m_3)$ was introduced. In
the coordinates $\{w_i\}$ the
constant in the conformal factor (\ref{confac}) is $\alpha=8$.
 
\section{Sectional curvatures}
 
In this section we study the eigenvalue problem of the Ricci tensor 
(\ref{Ricci}):
\begin{equation}
R_{\ j}^{i}\mathop{n}\limits_{\alpha }{}^{j}=\mathop{\rho}\limits_{\alpha }{}
\mathop{n}\limits_{\alpha }{}^{i}\ .  \label{eigen}
\end{equation}
The eigenvalues $\mathop{\rho}\limits_{\alpha }{}$ are $0$ and a double root
equal to $R/2$. All eigenvalues are well behaved for the configuration of
particles in a line. The conformal Killing vector $K^{i}$ (\ref{Ku}) is the
eigenvector corresponding to the zero eigenvalue. A two-dimensional subspace
spanned by the vectors:
\begin{eqnarray}
X_{(1)}^{i}=\left( 0,\ c/m_{3},\ -b/m_{2}\right) 
\nonumber\\ 
X_{(2)}^{i}=\left( -c/m_{3},\ 0,\ a/m_{1}\right)
\nonumber\\ 
X_{(3)}^{i}=\left( b/m_{2},\ -a/m_{1},0\right)
\end{eqnarray}
corresponds to the degenerate eigenvalue $R/2$. Any two of these vectors are
linearly independent, but not orthogonal to each other. From the general
theory of symmetric matrices we know that the two-dimensional space spanned
by $X_{(\nu )}^{i}$ is orthogonal to $K^{i}$. A compact notation for the
eigenvectors is:
\begin{equation}
K^{i}=a_{i}
\ ,\qquad 
X_{(\nu )}^{i}=\sum_{j}\epsilon_{\nu ij}{\frac{a_j}{m_j}}\
\end{equation}
and for a dual set of the eigenvectors:
\begin{eqnarray}
K_{i}={\frac{1}{2M}}\sum_{j,k}\mu _{ijk}a_{i}m_{j}m_{k}
\nonumber\\ 
Y_{i}^{(\nu)}=-\sum_{j}\mu_{\nu ij}m_{j}a_{i}a_{\nu }
+\delta _{i}^{\nu}\sum_{jk}\mu_{ijk}m_{j}a_{k}^{2}
\ .  \label{KY}
\end{eqnarray}
Each of the one-forms $Y_{i}^{(\nu )}$ is the dual of some vector $Y_{(\nu
)}^{i}$, orthogonal\footnote{
The vectors $Y_{(\nu )}^{i}$ themselves, however have cumbersome
expressions, therefore we will avoid to write them. They can be obtained by
the Gram-Schmidt orthogonalization procedure. The vectors $Y_{(\nu )}^{i}$
are not orthogonal to each other either.} 
to both $K^{i}$ and $X_{(\nu )}^{i}$.
 
It is immediate to check that the set of vectors $Y_{(\nu )}^{i}$ and $K^{i}$
are surface orthogonal, as they satisfy the relations:
\begin{equation}
K_{i,j}K_{k}\epsilon ^{ijk}=0\ ,\qquad Y_{i,j}^{(\nu )}Y_{k}^{(\nu
)}\epsilon ^{ijk}=0\ .
\end{equation}
We proceed to find these privileged surfaces. It follows 
from (\ref{Kd}) that the ellipsoids
\begin{equation}
F\equiv W-const=0\ ,
\end{equation}
are the surfaces orthogonal to $K^{i}$. Then the surfaces $f^{(\nu )}=0$ 
orthogonal to the vectors $Y_{(\nu )}^{i}$ can be determined by imposing the
proportionality of their gradients with $Y_{i}^{(\nu )}$: 
\begin{equation}
Y_{i}^{(\nu )}=\lambda ^{(\nu )}\nabla _{i}f^{(\nu )}\ ,  \label{Y}
\end{equation}
by some undetermined factor $\lambda ^{(\nu )}$. 
The detailed derivation is given below for the surface orthogonal to 
$Y_{(3)}^{i}$. 
We look for this surface in the form
\begin{equation}
f^{(3)}\equiv \Theta (c)-\Phi (a,b)\ .
\end{equation}
Comparing the expressions of $Y_{i}^{(3)}$ from (\ref{KY}) and (\ref{Y}) we
find three equations to determine the unknown functions $\Theta (c),$ $\Phi
(a,b)$ and $\lambda ^{(3)}$, which are:
\begin{eqnarray}
\lambda ^{(3)}\Theta _{,3}  = m_{1}b^{2}+m_{2}a^{2}  \nonumber \\
\lambda ^{(3)}\Phi _{,1}  = m_{2}ac  \label{eqs} \\
\lambda ^{(3)}\Phi _{,2}  = m_{1}bc\ ,  \nonumber
\end{eqnarray}
By inserting $\lambda ^{(3)}$ expressed from the first equation into the
rest of the system (\ref{eqs}), the remaining equations can be brought into
the form
\begin{equation}
\frac{c}{2}\frac{d\Theta }{dc}=\alpha =(m_{1}b^{2}+m_{2}a^{2})\frac{d\Phi }{
d\left( m_{1}b^{2}+m_{2}a^{2}\right) }\ ,
\end{equation}
which has the immediate solution $\Theta =\alpha \ln (\gamma c^{2})$ and 
$\Phi =\alpha \ln [\beta (m_{1}b^{2}+m_{2}a^{2})].$ Here $\alpha ,\
\beta $ and $\gamma $ are constants. Thus the equation of the surface
orthogonal to the vector $Y_{(3)}^{i}$ is given by:
\begin{equation}
f^{(3)}\equiv 
\alpha \ln\Bigl[\frac{\gamma c^2}{\beta (m_{1}b^{2}+m_{2}a^{2})}\Bigr]=0
\ ,
\end{equation}
or equivalently as
\begin{equation}
F^{(3)}\equiv m_{1}b^{2}+m_{2}a^{2}-const\cdot c^{2}=0\ ,
\end{equation}
Similar equations with cyclically permuted quantities define the surfaces to
which the other two vectors $Y_{(\nu )}^{i}$ are orthogonal. These surfaces
are cones with the tips in the origin and they have elliptical sections.
 
Next we derive the sectional curvatures of the privileged
sections $F$ and $F^{(\nu )}$. We pick up an independent set of eigenvectors
$\mathop{n}\limits_{\alpha }{}^{i}$ by normalizing $K^{i}$ and any two of
the three vectors $Y_{(3)}^{i}$. For each of these surfaces we define the
Riemann tensors $^{(2)}\hskip-.1cm\mathop{R}\limits_{\alpha }{}_{\ bcd}^{a}$
like in Eq. (46) of {\bf I}. We also define the projector operators $
\mathop{P}\limits_{\alpha }{}_{b}^{a}$ and the extrinsic curvatures $
\mathop{K}\limits_{\alpha }{}_{ab}$ for the sections:
\begin{equation}
\mathop{P}\limits_{\alpha }{}_{b}^{a}=\delta _{b}^{a}-\mathop{n}
\limits_{\alpha }{}^{a}\mathop{n}\limits_{\alpha }{}_{b}\ ,\qquad \mathop{K}
\limits_{\alpha }{}_{ab}=P_{a}^{c}P_{b}^{d}\nabla _{c}\mathop{n}
\limits_{\alpha }{}_{d}\ .
\end{equation}
The Gauss equation connects the curvature tensor and extrinsic curvature of 
the sections to the curvature tensor of the space of distances:
\begin{equation}
^{(2)}\hskip-.1cm\mathop{R}\limits_{\alpha }{}_{abcd}=\mathop{P}
\limits_{\alpha }{}_{a}^{e}\mathop{P}\limits_{\alpha }{}_{b}^{f}\mathop{P}
\limits_{\alpha }{}_{c}^{g}\mathop{P}\limits_{\alpha }{}_{d}^{h}R_{efgh}+2
\mathop{K}\limits_{\alpha }{}_{a[c}\mathop{K}\limits_{\alpha }{}_{d]b}\ .
\end{equation}
A double contraction with the induced metric on the sections 
$G^{ab}-\mathop{n}\limits_{\alpha }{}^{a}\mathop{n}\limits_{\alpha }{}^{b}$ 
gives the equations for the sectional curvatures:
\begin{equation}
^{(2)}\hskip-.1cm\mathop{R}\limits_{\alpha }{}=R-2\mathop{\rho}
\limits_{\alpha }{}+(\mathop{K}\limits_{\alpha }{}_{\ a}^{a})^{2}-\mathop{K}
\limits_{\alpha }{}_{ab}\mathop{K}\limits_{\alpha }{}^{ab}\ .  \label{scurv}
\end{equation}
Here we have employed that $\mathop{n}\limits_{\alpha }{}^{a}$ is
eigenvector of the Ricci tensor, Eq. (\ref{eigen}).
 
Straightforward computation based on the above prescription has shown that
the sectional curvature of the conical surfaces $F^{(\nu )}=0$ vanishes.
 
For the ellipsoid surfaces with the normal $n^{i}=K^{i}/\sqrt{2W}$ the
extrinsic curvature is found readily from the equation of the homothetic
motion (\ref{Killing}). In terms of $n^{i}$ the covariant form of this
equation is:
\begin{equation}
\nabla _{i}n_{j}+\nabla _{j}n_{i}=\sqrt{\frac{2}{W}}\left(
G_{ij}-n_{i}n_{j}\right) \ .  \label{homocov}
\end{equation}
Contracting twice with the projectors to the ellipsoid surface, the desired
result emerges:
\begin{equation}
K_{ij}={\frac{1}{\sqrt{2W}}}\left( G_{ij}-n_{i}n_{j}\right) \ .
\end{equation}
A simple computation then yields
\begin{equation}
(\mathop{K}\limits_{\alpha }{}_{\ a}^{a})^{2}-\mathop{K}\limits_{\alpha
}{}_{ab}\mathop{K}\limits_{\alpha }{}^{ab}={\frac{1}{W}}
\end{equation}
and from (\ref{scurv}), the sectional curvature of the ellipsoid surface $
F=0 $ is found:
\begin{equation}
{}^{(2)}R={\frac{4}{W}}={\frac{4}{3}}R\ .  \label{scurvellips}
\end{equation}
This is constant on the ellipsoids, therefore in the metric $G_{ij}$ the $
F=0 $ surfaces are spheres.
 
\section{Geodesic motions}
 
The equation for the homothetic motion (\ref{Killing}) carries even more
information. A contraction of its covariant form with $K^{j}$ gives:
\begin{equation}
K^{j}\nabla _{j}K_{i}+{\frac{1}{2}}(K_{j}K^{j})_{,i}=2K_{i}\ .
\end{equation}
The second term is simply $K_{i}$ as can be seen from (\ref{Kd}). Thus the
integral curves of the conformal Killing vector field $K^{i}$ are geodesics:
\begin{equation}
K^{j}\nabla _{j}K_{i}=K_{i}\ .  \label{geodK}
\end{equation}
We recall that $K^{i}\partial /\partial x^{i}=\partial /\partial \ln r.$
Therefore $\ln r$ is not an affine parameter, however it is easy to check
that $pr+q$ is, if $p$ and $q$ are constants.
 
Eq. (\ref{geodK}) says that the motions along radial lines of the reduced
space are geodesics. Therefore the boundaries of the reduced space (Fig. 1)
are geodesic planes in the sense that they are spanned by geodesics passing
through the origin. The privileged conical surfaces $F^{(\nu )}=0$ are
geodesic surfaces in the same sense.
 
The ellipsoid sections $F=0$ are not geodesic surfaces. To see this, we
compute the Gaussian curvature ${\cal K}$ of a geodesic surface whose
tangent space in each point is spanned by any two of the (non-orthogonal)
eigenvectors $Y_{(\mu )}^{i}$ by means of \cite{Eisenhart}
\begin{equation}
\ {\cal K}=\frac{R_{hijk}Y_{(\mu )}^{h}Y_{(\nu )}^{i}Y_{(\mu )}^{j}Y_{(\nu
)}^{k}}{\left( G_{nl}G_{mp}-G_{lp}G_{mn}\right) Y_{(\mu )}^{l}Y_{(\nu
)}^{m}Y_{(\mu )}^{n}Y_{(\nu )}^{p}}\ .
\end{equation}
We have found $\ {\cal K}=R/2$. Since the Gaussian curvature is half of the
curvature scalar, the latter would be $^{(2)}R_{geod}=2{\cal K}=R$ for a
geodesic surface with tangents $Y_{(\mu )}^{i}$. This is different from the
scalar curvature (\ref{scurvellips}) of the ellipsoid sections with the same
tangents. Therefore the ellipsoid sections cannot be geodesic surfaces (in
each point they have common tangents with {\it different} geodesic surfaces).
 
In order to have the generic geodesic motions
\begin{equation}
\frac{d^{2}a_{i}}{d\lambda ^{2}}+\Gamma _{jk}^{i}\frac{da_{j}}{d\lambda }
\frac{da_{k}}{d\lambda }=0\ ,  \label{geodeq}
\end{equation}
we need the Christoffel symbols $\Gamma _{jk}^{i}$ of the space of orbits.
They are given as
\begin{eqnarray}
\Gamma _{jk}^{i}  = \frac{m_{1}m_{2}m_{3}}{16M^{3}W^{2}Z^{2}}m_{i}^{2}
\Biggl\{ \mu _{ijk}a_{i}a_{j}a_{k}\Lambda _{ijk}\Lambda _{ikj}
\nonumber \\
-\delta_{jk}a_{i}a_{j}^{2}\sum_{p}\mu _{ijp}\Lambda _{ipj}^{2}  
-\delta_{jk}\delta _{ij}\frac{1}{a_{i}}\Bigl( \sum_{p,q}\epsilon
_{ipq}\Xi _{pq}\Bigr) ^{2}
\nonumber \\
+\left[ \delta _{ik}a_{j}\sum_{p}\mu
_{ijp}\Lambda _{ipj}(\Xi _{pj}-\Xi _{jp})+(j\leftrightarrow k)\right]
\Biggr\}\ .
\end{eqnarray}
In the above formula and the forthcoming ones all summations are indicated
explicitly. The coefficients $\Lambda _{ijk}$ and $\Xi _{ij}$ have the
expressions:
\begin{eqnarray}
\Lambda _{ijk}  = \Bigl( \sum_{r}a_{r}^{2}-2a_{i}^{2}\Bigr)
m_{j}+2a_{j}^{2}m_{k}  \label{f1} \\
\Xi _{ij}  = \Bigl( \sum_{r}a_{r}^{2}-2a_{i}^{2}\Bigr) a_{i}^{2}m_{j}
\label{f2}
\end{eqnarray}
We note here that in the $\{a_i\}$ coordinates $\Gamma _{jk}^{i}$ diverges 
on the boundaries.  
The geodesic equations (\ref{geodeq}), supplemented by the boundary
conditions (\ref{refl1}) and (\ref{refl2}) represent the
equations of motion for the three-body problem with vanishing constants of
motion, constant potential case, in terms of relative separations.
 
We seek for solutions of the geodesic equation in the following form
\footnote{
For generic potentials $V$ we do not dispose of such a solving
Ansatz, as is
well known from the three-body problem \cite{Whittaker}. }
\begin{equation}
\lambda =t\ ,\qquad a_{i}(t)=\sqrt{A_{i}t^{2}+B_{i}t+C_{i}}\ .
\label{straightline_uniform}
\end{equation}
These characterize the uniform velocity, straight line motions of the
particles in the physical space. For such motions the vanishing of the
energy is assured by a proper choice of the constant $V$. The coefficients $
A_{i},B_{i}$ and $C_{i}$ are functions of the relative positions ${\bf r}
_{jk}={\bf r}_{j}-{\bf r}_{k}$ and relative velocities ${\bf \dot{r}}_{jk}$
of the three particles at some initial time:
 
\begin{equation}
A_{i}\!=\!\frac{1}{2}\sum_{j,k}\!\mu _{ijk}{\bf \dot{r}}_{jk}^{2}\ ,\quad
B_{i}\!=\!\sum_{j,k}\mu _{ijk}{\bf r}_{jk}\!\cdot\! {\bf \dot{r}}_{jk}\ ,\quad
C_{i}\!=\!\frac{1}{2}\sum_{j,k}\!\mu _{ijk}{\bf r}_{jk}^{2}\ .  \label{ABCdef}
\end{equation}
Therefore $A_{i}=0$ and $C_{i}=0$ both imply $B_{i}=0$ too.
 
It is straightforward to check that particular cases of the motions (\ref
{straightline_uniform}) fulfill the geodesic equation (\ref{geodeq}). In the
case $A_{i}=0$ the geodesic equations are trivially satisfied, this choice
of the constants corresponding to no motion at all in the space of orbits.
When $C_{i}=0$, then $a_{i}(t)=\sqrt{A_{i}}t$ are geodesics passing through
the origin, with affine parameter $t$. These are the motions with tangents $
K^{i}$, the affine parameter $t$ being linearly related to the radial
coordinate $r$. The same type of motions $a_{i}(t)=\sqrt{A_{i}}t+\sqrt{C_{i}}
$ emerge also for the choice $B_{i}=2\sqrt{A_{i}C_{i}}$. In this latter case
Eqs. (\ref{ABCdef}) imply $A_{i}=\varphi C_{i}$ therefore ${\bf \dot{r}}
_{jk}=\varphi {\bf r}_{jk}$. The special case of a pure expansion motion (a
similarity transformation of the triangle of Fig. 2) is included here as the
particular case with ${\bf \dot{r}}_{i}=\varphi {\bf r}_{i}$.
 
Those motions (\ref{straightline_uniform}) which do not violate the
constraints ${\bf P}=0$ and ${\bf L}=0$ are expected to be geodesics with
affine parameter $t$. The fact that not all motions (\ref
{straightline_uniform}) are geodesics is readily seen for the case $B_{i}=0$
with all other constants nonvanishing. In this case the change of the
relative velocities{\bf \ }${\bf \dot{r}}_{jk}$ is orthogonal to the
relative positions ${\bf r}_{jk}$. When all $A_i$ are equal, this situation 
corresponds to an overall rotation of the system.
 
The integral form of the generic geodesics is found as follows. Considering 
a motion of the type (\ref{straightline_uniform}) through a point $a_{i}(0)$
with direction $\dot{a}_{i}(0)$ at $t=0$, we find that the constants $B_{i}$
and $C_{i}$ are determined directly by the initial data
\begin{equation}
B_{i}=2a_{i}(0)\dot{a}_{i}(0)\ ,\qquad C_{i}=a_{i}(0)^{2}\ .  \label{BiCi}
\end{equation}
By enforcing the motions (\ref{straightline_uniform}) to be geodesic 
through Eq. (\ref{geodeq}), three
polynomial identities in $t$ are found, each of degree 6. As the geodesic
condition holds at all instants, we can impose that all coefficients
vanish. It turns out that the coefficients of $t^{0}$ contain the
constants $A_{i}$ linearly. Therefore, (or equivalently from the
geodesic condition at $t=0$) the constants $A_{i}$ can be
expressed in terms of $B_{i}$ and $C_{i}$. It is then verified than with
these $A_{i}$ all other coefficients are also zero. The result
of the above computation is
\begin{eqnarray}
  A_{i} \!=\! \Biggl\{ -m_{i}^{2}C_{i}
\Bigl( \sum_{j,k}\!\epsilon _{ijk}\Lambda_{ijk}B_{k}\Bigr)^{2}
+2m_{i}^{2}B_{i}
\Bigl(\sum_{j,k}\!\epsilon_{ijk}\Lambda _{ijk}B_{k}\Bigr) 
\Bigl(\sum_{n,l}\!\epsilon _{inl}\Xi_{nl}\Bigr)   
\nonumber \\
  +B_{i}^{2}\Biggl[
\sum_{j,k}\mu _{ijk}m_{j}m_{k}
\Bigl(\gamma^{(3)}-2\gamma^{(2)}C_{i}+\gamma^{(1)}C_{i}^{2}
+\frac{\Xi_{ij}\Xi_{ik}}{2C_{i}}\Bigr) 
\nonumber \\
  +\Bigl(m_1m_2m_3\sum_{l}\!C_{l}m_{l}-2\gamma^{(0)}\gamma^{(1)}\Bigr) 
\sum_{j,k}\!\mu _{ijk}C_{j}C_{k}  
\nonumber\\
  +2m_{1}m_{2}m_{3}m_{i}C_{1}C_{2}C_{3}
\Biggr ]\Biggr\}  
\nonumber \\
  /4\left( \gamma ^{\left( 1\right) }\right) ^{2}\left(
C_{1}^{2}+C_{2}^{2}+C_{3}^{2}-2C_{1}C_{2}-2C_{2}C_{3}-2C_{3}C_{1}\right) \ ,
\label{Ai}
\end{eqnarray}
where $\Lambda _{ijk}$ and $\Xi _{jk}$ are given by Eqs. (\ref{f1}), (\ref
{f2}) and $\gamma ^{\left( n\right) }$ by
\begin{equation}
\gamma ^{\left( n\right)
}=C_{1}^{n}m_{2}m_{3}+C_{2}^{n}m_{3}m_{1}+C_{3}^{n}m_{1}m_{2}\ .  \label{f3}
\end{equation}
The quantity $\gamma ^{\left( 1\right) }$ is just $2MW$ introduced
before. By
taking the common denominator of Eq. (\ref{Ai}), we find sums of expressions
of the type $A_{i}C_{j}^{4}$ and $B_{k}^{2}C_{l}^{3}$ on the left and right
hand sides, respectively. It is easy to check, that all previously discussed
special cases apply.

The expressions (\ref{Ai}) are indeterminate on the boundaries. 
Indeed, it can be shown that there the numerators are proportional to 
${\bf \dot a}_\perp$. The presence of the vanishing
factor in the denominator is related to the singular behavior of 
$\Gamma _{jk}^{i}$ in the collinear configurations. 

Nevertheless, by starting the motion from any noncollinear configuration, 
the coefficients $A_i$ are well defined, and  
Eq. (\ref{straightline_uniform}) together with the coefficients (\ref{BiCi})
and (\ref{Ai}) represent the general solution of the geodesic equation in
the constant potential case. 
With this, the Cauchy problem is also solved:
we have found the motions pertinent to arbitrary noncollinear initial data 
$\left( a_{i},\dot{a}_{i}\right)$.
 
\section{Concluding Remarks}
 
We have presented a detailed analysis of the geometry of the space of orbits 
of the Leibniz group for three particles. We have shown that in the line 
configurations (corresponding to the boundaries of the reduced space) no 
curvature singularity occurs. The boundary conditions (\ref{refl1}) and 
(\ref{refl2}) imply that the geodesics reach the boundaries 
tangentially after which they must return to the inner region.

Our geometrical approach provides an alternative to the classical treatments 
of the three-body problem \cite{Whittaker}. As has been shown first by 
Lagrange, a generic reduction of the $18^{th}$ order system of differential 
equations characterizing the three-body problem to a $6^{th}$ order system 
is guaranteed by the existence of the ten integrals of motion. For generic 
situations no other reduction can be made, as the theorem of Bruns forbids 
the existence of any further (algebraically independent) integral of motion. 
The $6^{th}$ order system characterizing the particular motions with 
vanishing first integrals is the set of geodesic equations (\ref{geodeq}). 
Supplemented by the boundary conditions (\ref{refl1}) 
and (\ref{refl2}), they describe free motions.
We have investigated the constant potential case, however generalization is 
straightforward by a conformal rescaling of the metric, with the conformal 
factor $-2V$. We hope that the geometric methods developed in this paper 
will be applied in the study of nontrivial dynamical problems either, 
pertinent to specific $V\neq const$ cases.

For completeness we compute the Coriolis tensor for three particles \cite{LR} 
which was introduced initially in the context of molecular dynamics. 
In a body frame differing from the principal axis frame by a rotation about 
the third axis and choosing the basis $da_i$ in the shape space it has the 
components: 
\begin{equation}
B^{\nu}_{bc}=-\delta^{\nu}_3
             \Bigl(\frac{m_1m_2m_3}{2M}\Bigr)^{3/2}
             \frac{abc}{ZW^2}
             \epsilon_{bca}K^a\ .
\label{Coriolis}
\end{equation}
Here the index $\nu$ refers to the body frame and indices $a,\ b$ and $c$ to 
the shape space (our space of distances).
We would like to stress that the Coriolis tensor (\ref{Coriolis}) is related in 
a simple way to the conformal Killing vector $K^a$, 
the existence of which we have exploited in many ways in this paper.
 
\ack
 
The authors are grateful to Karel Kucha\v {r} for fruitful interactions 
during the elaboration of this work. The criticism of the referees on a 
previous version of the boundary conditions is acknowledged, as it led
to substantial improvements.
L.\'A.G. was supported by the NSF grant PHY-9734871, OTKA grants
W015087 and D23744, the E\"otv\"os Fellowship and the Soros Foundation.

\end{document}